\documentstyle[pra,aps]{revtex}
\tighten
\begin{document}
\preprint{}
\draft
%
%%%%%%%%%%%%%%%%%%%%%%%%%%%%%%%%% TITLE PAGE
%
\title{Self-consistent model for ambipolar tunneling\\
in quantum-well systems}
\author{C. Presilla, V. Emiliani and A. Frova}
\address{Dipartimento di Fisica, 
Universit\`a di Roma ``La Sapienza'',\\
Piazzale A. Moro 2, Roma, Italy 00185}
\date{Semiconductors Science \& Technology {\bf 10} (1995) 577-585}
\maketitle
%
%%%%%%%%%%%%%%%%%%%%%%%%%%%%%%%%% ABSTRACT
%
\begin{abstract}
        
We present a self-consistent approach to describe ambipolar tunneling 
in asymmetrical double quantum wells under steady-state excitation and 
extend the results to the case of tunneling from a near-surface quantum 
well to surface states. 
The results of the model compare very well with the behavior observed in 
photoluminescence experiments in $InGaAs/InP$ asymmetric double quantum wells
and in near-surface $AlGaAs/GaAs$ single quantum wells.
                                                              
\end{abstract}
%
%%%%%%%%%%%%%%%%%%%%%%%%%%%%%%%%% PACS NUMBERS
%
%
%%%%%%%%%%%%%%%%%%%%%%%%%%%%%%%%% PAPER BODY
%
\section{Introduction}

        Tunneling of electrons and holes in quasi-two-dimensional semiconductor
heterostructures is the object of large interest both for its physical interest
- it is one of the most important quantum-mechanical effect observed in 
low-dimensional structures 
- and for its role in several nanometric devices and applications.
Much work has been done concerning
tunneling in symmetrical and asymmetrical double quantum wells (ADQW).

Tunneling from a quantum well (QW) to surface states has received 
instead little 
attention, despite the great importance of gaining control over this mechanism.
For a quantum well built in the neighborhood of
an unpassivated surface, an extra non-radiative recombination channel becomes
available to electrons and holes if they can tunnel to surface states, with 
consequent loss 
in emission efficiency from the quantum well. The importance of this effect 
has been demonstrated experimentally in 
recent papers and its dependence on surface-barrier thickness investigated
\cite{FE,2,3,4}.

In the interpretation of the experimental results, both tunneling 
to surface states and tunneling between ADQW, a problem arises.
Although in
some cases hole-tunneling rates can be comparable to the electronic ones, e.g. 
when
heavy holes in one well move to a light hole state in the other well \cite{5}, 
direct-gap III-V
semiconductors have in general rather different tunneling probabilities for 
electrons and holes
due to the different effective masses \cite{6,7}. The different tunneling 
probability for the two
carriers causes a dipole electric field across the tunneling barrier to 
develop \cite{japl,STT,9,10},
so as to induce - via the quantum confined Stark effect \cite{bea}
- a peak shift of the excitonic 
recombination, and to affect the tunneling probabilities towards an 
ambipolar regime, with equal
tunneling currents for electrons and holes. 

The appearance of an electric field due to the spatial separation 
of electrons and holes in ADQW systems was already discussed theoretically
\cite{FERR} in the framework of exciton tunneling under impulsive
excitation. 
In this paper we study the tunneling of unpaired electrons and holes 
both in ADQW and in near-surface QW systems under steady-state excitation. 
Unlike the case of impulsive excitation, 
not discussed in this paper, the ambipolar regime is shown 
to be reached in this case for any excitation intensity.
The consequently modified tunneling properties are successfully
compared with available experiments, i.e. dependence of
the emission efficiency ratio in 
ADQW systems on the excitation power and the tunneling-barrier width 
\cite{STT}, 
dependence of the
tunneling current on the barrier thickness \cite{FE}, and
dependence of the 
Stark shift on the excitation
level in a near-surface QW \cite{japl}.

\section{tunneling between two asymmetric quantum wells}
                          
Here we describe the steady-state photoluminescence from two 
coupled asymmetric quantum wells under constant irradiation.
The charges photogenerated inside the wells relax almost instantaneously 
with respect to the other relevant time scales to the lowest band of 
the respective well and fill it according to the Pauli principle. 
Electron-hole interaction leads to exciton formation.
Tunneling between the two wells is essentially restricted to unpaired 
electrons and holes when the tunneling process is a small perturbation
for the nearly uncoupled wells: 
as a matter of fact, the tunneling current densities correspond to a
first-order process for unpaired electrons and holes and to a 
second-order process for excitons.  
On the other hand photoluminescence is restricted only to excitonic 
recombination.
                                      
Let $a_1$ and $a_2$ be the widths of the two quantum wells 
and $b$ the width of the barrier between them.
We suppose $a_1 > a_2$ so that the bottom of the $e1$ and $hh1$ bands of 
well 2, $E_{e1}^{(2)}$ and $E_{hh1}^{(2)}$, are higher in energy than 
those of well 1, $E_{e1}^{(1)}$ and $E_{hh1}^{(1)}$.
Let $G_1$ and $G_2$ be the generation current densities of 
electron-hole pairs in the two wells.
If $n_i$ and $p_i$ are the steady-state concentrations 
(number of particles per unit area) of electrons and holes 
in the well $i=1,2$, the following rate equations hold:
\begin{mathletters}
\label{RE}
\begin{equation}
0 = G_1 + J_{e} - \lambda_1 n_1 p_1
\end{equation}
\begin{equation}
0 = G_1 + J_{h} - \lambda_1 n_1 p_1
\end{equation}
\begin{equation}
0 = G_2 - J_{e} - \lambda_2 n_2 p_2
\end{equation}
\begin{equation}
0 = G_2 - J_{h} - \lambda_2 n_2 p_2
\end{equation}
\begin{equation}
0 = \lambda_1 n_1 p_1  - I_{1}
\end{equation}
\begin{equation}
0 = \lambda_2 n_2 p_2  - I_{2}    .
\end{equation}
\end{mathletters}

The first two couples of equations are the rate equations for 
the concentrations of unpaired electrons and holes in well 1 and 2, 
respectively, and contain, in the order, the photogeneration and 
tunneling current densities of the unpaired charges and the generation 
current density of excitons. 
The last two equations are the rate equations for the exciton 
concentrations in well 1 and 2 and contain the photoluminescence current 
densities $I_1$ and $I_2$ which are the quantities we want to evaluate
and compare with experimental results.

The generation current density of excitons in well $i$ is assumed 
proportional to the concentrations of the unpaired charges in the same well,
$\lambda_i n_i p_i $, where $\lambda_i$ is the bimolecular exciton 
formation coefficient possibly dependent on the well size \cite{SEKK}.

A little more complicated analysis is needed for the tunneling
current densities $J_{e}$ and $J_{h}$.
Transfer of electrons (holes) from the narrower well to the larger one 
is realized in a non-coherent two-step process. 
Quantum coherent tunneling of electrons (holes) from an occupied state of
the $e1$ ($hh1$) band of well 2 to an equal-energy empty state of 
the $e1$ ($hh1$) band of well 1 is followed by thermalization 
toward the lower energy states of well 1.
When the barrier width $b$ is not too small, which is also the range 
of validity of the above rate equations, the quantum coherent 
tunneling process characterized by a time growing exponentially
with $b$ gets much slower than the thermalization process 
in well 1 ($\lesssim$ 1 ps \cite{NAGARA}) and this last can be neglected.

The current densities for quantum coherent tunneling are approximately 
proportional to the charge concentrations in well 2 and the proportionality 
factor, namely the tunneling rate, is quite different for electrons and 
holes due to their different effective masses.
Therefore, in a steady-state situation 
when $J_{e}=J_{h}$, the concentrations of electrons and holes
in each single well must be different and an electric field 
gets established in the barrier between the two wells.
Moreover the electric field modifies the electron and hole tunneling
rates.
The direction of the field is simply understood from the values 
of the tunneling rates at zero field.
As electron tunneling rate is expected in this case to be much
greater than the hole tunneling rate, electrons accumulate at well
1 and in a steady-state situation $n_1>p_1$.
The electric field is directed from well 2 to well 1 and its value
is given by
\begin{equation}
F={en \over \varepsilon_0 \varepsilon_r} 
\label{F}
\end{equation}
where $n=n_1-p_1=p_2-n_2$ and $\varepsilon_r$ is the permittivity
of the barrier material. 

The tunneling current densities depend on the electric field $F$ 
through the dependence of the charge concentrations on $F$
as well as through the dependence of the tunneling rate on the energy shift 
between the bands of the two wells induced by $F$.
Assuming that the electric field is completely shielded inside the wells,
this energy shift amounts to $eFb$.
For a given value of $F$, the tunneling current densities 
can be evaluated within perturbation theory in the case of a 
barrier not too thin.  
At the first order the transition rate induced by a constant perturbation $V$ 
between initial and final states continuously distributed in energy 
with densities $d N_i / d \epsilon$ and $d N_f / d \epsilon$, respectively, 
is \cite{LANDAU}
\begin{equation}
\int \int {2 \pi \over \hbar} 
\left| \langle i,\epsilon | V | f,\epsilon' \rangle \right|^2  
\delta(\epsilon -\epsilon') d N_f(\epsilon') d N_i(\epsilon) 
=
\int {2 \pi \over \hbar} 
\left| \langle i,\epsilon | V | f,\epsilon \rangle \right|^2  
{d N_f \over d \epsilon} (\epsilon) 
{d N_i \over d \epsilon} (\epsilon)  ~d \epsilon .
\end{equation}
This formula, when applied to the electron and hole states of the
two uncoupled wells and divided by the transverse area $A$ of the   
heterostructure, gives the following expressions for the tunneling 
current densities 
\begin{equation}
J_{e} = {1 \over A} \int_0^\infty {2 \pi \over \hbar} 
\left| \langle \Phi^{(1)}_{\epsilon} |
V_e| \Phi^{(2)}_{\epsilon} \rangle \right|^2  A~ \nu_e^{(1)}
\left[ 1-f\left({\epsilon - \epsilon_F^{(1)} \over k_B T} \right) \right]
A~ \nu_e^{(2)} f\left({\epsilon - \epsilon_F^{(2)} \over k_B T} \right)  
~d\epsilon 
\label{JEG}
\end{equation}
\begin{equation}
J_{h} = {1 \over A} \int_0^\infty {2 \pi \over \hbar} 
\left| \langle \Phi^{(1)}_{\epsilon} |
V_h| \Phi^{(2)}_{\epsilon} \rangle \right|^2  A~ \nu_h^{(1)} 
\left[ 1-f\left({\epsilon - \epsilon_F^{(1)} \over k_B T} \right) \right]
A~ \nu_h^{(2)} f\left({\epsilon - \epsilon_F^{(2)} \over k_B T} \right)
~d\epsilon  ~.
\label{JHG}
\end{equation}
The integration takes into account the transitions from all 
the occupied states in the $e1$ ($hh1$) band of well 2 (initial states) 
to the empty states in the $e1$ ($hh1$) band of well 1 (final states)
at the same energy $\epsilon$ measured 
from the bottom of $e1$ ($hh1$) band of well 2.
The densities of the available initial (final) states are obtained 
by multiplying the number of states per unit energy $A~ \nu^{(2)}$
($A~ \nu^{(1)}$) with the appropriate occupation probability. 
Here $\nu_e^{(i)}=m_e^{(i)}/\pi \hbar^2$ 
($\nu_h^{(i)}=m_h^{(i)}/\pi \hbar^2$), i=1,2,  are 
the densities per unit area and unit energy of a 
two dimensional ideal gas of fermions with mass $m_e^{(i)}$ ($m_h^{(i)}$)
and $f(x)=1/[\exp(x)+1]$ is the Fermi function.
The Fermi energies for electrons (holes) are related to the  
concentrations in the corresponding well: 
$\epsilon_F^{(1)}=\pi \hbar^2 n_1/ m_e^{(1)}+E_{e1}^{(1)}+eFb-E_{e1}^{(2)}$
and $\epsilon_F^{(2)}=\pi \hbar^2 n_2/m_e^{(2)}$
($\epsilon_F^{(1)}=\pi \hbar^2 p_1/m_h^{(1)}+E_{hh1}^{(1)}-eFb-E_{hh1}^{(2)}$
and $\epsilon_F^{(2)}=\pi \hbar^2 p_2/m_h^{(2)}$). 
Finally, the perturbation potential $V_e$ ($V_h$) 
is the potential of the heterostructure ($b$ finite)  
coupling the electron (hole) states evaluated by considering 
the wells uncoupled ($b$ infinite). 
Due to the exponentially vanishing tails of the electron (hole) wavefunctions,
the relevant contribution to the matrix elements comes only from
the barrier region between the two wells \cite{LANDAU}.
In this region the potential $V_e$ ($V_h$) has magnitude of the order 
of the conduction (valence) band offset $\Delta E_c$ ($\Delta E_v$).
More accurate expressions for $V_e$ ($V_h$), e.g. taking into account
the distortion due to the electric field, have little influence 
on the final result and will be neglected.
Explicit expressions of the electron and hole states $\Phi$
and analytical evaluation of the matrix elements are given in Appendix A.

The formulas for the tunneling current densities can be simplified 
by noting that, due to the exponentially large difference between
electron and hole tunneling rates, we expect $n_2 \ll n_1$.
As a consequence, the electron Fermi energy $\epsilon_F^{(2)}$, 
i.e. the effective region of integration in Eq.\ (\ref{JEG}), is very small. 
By approximating all factors to be integrated in (\ref{JEG}) but the 
Fermi function in well 2 with their value at $\epsilon=0$
(the bottom of the $e1$ band of well 2) we get $J_{e}=n_2/\tau_e$, where 
$n_2=\int \nu_e^{(2)} f\left((\epsilon - \epsilon_F^{(2)} )/ k_B T \right)  
d\epsilon$ and 
\begin{equation}
{1 \over \tau_e} = {2 \pi \over \hbar} 
\left| \langle \Phi^{(1)}_0 | V_e| \Phi^{(2)}_0 \rangle \right|^2   
A~\nu_e^{(1)}
f\left({ \epsilon_F^{(1)} \over k_B T} \right) ~.
\label{TAUE}
\end{equation}
In the case of holes we expect $p_2 \gg p_1$ so that $\epsilon_F^{(2)}$ 
is not small. 
However, in this case $\epsilon_F^{(1)}<0$ (now the zero of energy is
the bottom of the $hh1$ band of well 2) and
$f((\epsilon - \epsilon_F^{(1)})/k_B T) \simeq 0$.
Neglecting the smooth dependence of the matrix element on $\epsilon$
in (\ref{JHG}) we get $J_{h}=p_2/\tau_h$ where 
$p_2=\int \nu_h^{(2)} f\left((\epsilon - \epsilon_F^{(2)} )/ k_B T \right)  
d\epsilon$ and 
\begin{equation}
{1 \over \tau_h} = {2 \pi \over \hbar} 
\left| \langle \Phi^{(1)}_0 |V_h| \Phi^{(2)}_0 \rangle \right|^2 
A~ \nu_h^{(1)} ~. 
\label{TAUH}
\end{equation}
Note that both the tunneling rates $1/\tau_e$ and $1/\tau_h$ depend
on the electric field through the matrix elements (see Appendix A).
In addition the electron tunneling rate depends on $F$ through the 
Fermi energy of the well 1.

If we think that the electric field, i.e. $n$, and the 
tunneling rates are known, we get the following solution 
for the initial set of rate equations
\begin{mathletters}
\label{SORE}
\begin{equation}
n_2 = \sqrt{ \left({n\over2}+{1\over2\lambda_2 \tau_e} \right)^2 +
{G_2 \over \lambda_2} } -
\left({n\over2}+{1\over2\lambda_2 \tau_e} \right)
\end{equation}
\begin{equation}
n_1 = \sqrt{ \left({n\over2} \right)^2 +
{G_1 \over \lambda_1} + {n_2 \over \lambda_1 \tau_e} } + {n\over2}
\end{equation}
\begin{equation}
p_1=n_1-n
\end{equation}
\begin{equation}
p_2=n_2+n ~.
\end{equation}
\end{mathletters}
This result allows us to find the steady-state values of the 
electric field, of the charge concentrations and of the tunneling rates 
by a recursive method.
We proceed in the following manner.
First of all we fix an electric field value
corresponding to some negative electric charge concentration $n$
in the well 1.
Then we evaluate the hole tunneling rate and 
give a starting value to the electron tunneling rate (for instance 
that obtained for $n_1=n$) so we can deduce some charge concentrations
by Eq.\ (\ref{SORE}).
From these concentrations we evaluate a new electron tunneling rate 
by Eq.\ (\ref{TAUE}) and from that new charge concentrations.
After few iterations, the electron tunneling rate and the 
charge concentrations converge to the solution corresponding to
the value of the electric field fixed at the beginning.
At this point we compare the values of the tunneling current densities.
If $J_{e}>J_{h}$ the electric field is increased, 
if $J_{e}<J_{h}$ the electric field is decreased. 
The procedure is repeated until the steady-state condition
$J_{e}=J_{h}$ is reached. 
Note that the existence of the ambipolar regime does not rely upon
the value of the excitation intensity, i.e. $G_1$ and $G_2$.
However, the value of the electric field established in the steady 
state strongly depends on the excitation intensity.

We illustrate the behavior of the model by considering a practical
example close to the experimental situation investigated by
Sauer, Thonke and Tsang \cite{STT}.
We consider two $In_{0.53}Ga_{0.47}As$ quantum wells imbedded between 
$InP$ barriers with $a_1=100$ \AA\ and $a_2=60$ \AA.
We use the following material parameters at $T=4.2$ K \cite{LB}: 
$\Delta E_c=0.195$ eV, $\Delta E_v=0.293$ eV, 
$m_e^{(1)}=m_e^{(2)}=0.044~m$, $m_h^{(1)}=m_h^{(2)}=0.38~m$, 
$m$ being the free electron mass, and $\varepsilon_r=13.9$. 
Moreover we put $\lambda_1=\lambda_2=6$ cm$^2$ s$^{-1}$ \cite{SEKK}.
Since $\alpha a_1$, $\alpha a_2 \ll$ 1, where $\alpha $ is the optical 
absorption
coefficient for the pump light, we take $G_1/G_2=a_1/a_2$ and $G_2=P/h\nu$ 
where $P$ is the absorbed 
power density and $h\nu=2.41$ eV is the photon energy corresponding 
to a 514-nm laser light.
Note that the transverse area $A$ shown in Eq.\ (\ref{TAUE})
and Eq.\ (\ref{TAUH}) cancels out with the $A^{-1}$ from
the squared matrix elements and is an irrelevant parameter. 

In Fig. 1 we show the calculated photoluminescence intensity ratio $I_1/I_2$ 
from the two wells as a function of the absorbed power density $P$
for different values of the barrier width $b$.
The self-consistent electric field $F$ generated between the two wells 
in the same cases is shown in Fig. 2.
When $P$ is lower than a critical value, 
which depends on the barrier width, 
the photoluminescence intensity ratio $I_1/I_2$ approximately decreases 
as $P^{-1}$ and the electric field increases as $P$. 
When $P$ reaches and exceeds the critical value 
an electric field of the order of $3 \div 4 \times 10^4$ V cm$^{-1}$  
gets established which slows down the tunneling of electrons from well 
2 to well 1 and enhances the photoluminescence from well 2.
At very high absorbed power density the ratio $I_1/I_2$  
tends to the value $G_1/G_2$ for two uncoupled wells.
These results compare quite well with the experimental findings
\cite{STT} if one assumes a 0.1 \%  excitation efficiency. 

A deeper understanding of the behavior shown in Fig.s 1 and 2  
can be reached by an approximate solution of the self-consistent
procedure described above.
Let us first concentrate on the dependence of the electric field,
i.e. $n$, on the absorbed power density $P$.
In the steady state we have $n_2 / \tau_e = p_2 / \tau_h$
which combined with Eq.\ (\ref{SORE}a) and Eq.\ (\ref{SORE}d)
gives
\begin{equation}
n = {1 \over 2 \lambda_2} 
\left( {1\over \tau_e} - {1\over \tau_h} \right)
\left( 
\sqrt{ 1 + {4 \lambda_2 \tau_e \tau_h \over h \nu} ~P} - 1 
\right) .
\end{equation}
When $P$ is not too high the second term in the squared root 
is small and we get the simple self-consistent equation
\begin{equation}
n + {\tau_e(n) \over h \nu} ~P = {\tau_h \over h \nu} ~P .
\label{AS}
\end{equation}
In a first approximation the smooth dependence of the tunneling rates 
on the electric field through the corresponding matrix elements can be 
neglected so that $\tau_h$ is constant while $\tau_e$ still depends on 
$n$ through the Fermi function.
The argument of the Fermi function in Eq.\ (\ref{TAUE}) is negative until 
the alignment of the electron bands is achieved due to the electric field,
i.e. in the region $n \leq n_{crit}$ where
$e^2 b~n_{crit} / \epsilon_0 \epsilon_r = E_{e1}^{(2)} - E_{e1}^{(1)}$.
In this range $\tau_e$ is approximately constant and much smaller
than $\tau_h$.
On the other hand for $n > n_{crit}$, the electron tunneling time
increases exponentially with the adimensional parameter  
$eFb / k_B T$, i.e. with $n$.
The behavior of $\tau_e(n)$ imposes different solutions to 
Eq.\ (\ref{AS}) at different values of the absorbed power density.
At low $P$ we have $\tau_e(n) P / h \nu \ll n$ and therefore
$n \simeq \tau_h P / h \nu$.
At high $P$ we have $\tau_e(n) P / h \nu \gg n$ and therefore
$n \simeq n_{crit}$ is constant. 
The critical value of the electric field is related to the 
critical value of the absorbed power density by 
$P_{crit} \simeq h \nu n_{crit} / \tau_h$.
Since $\tau_h$ increases exponentially with the barrier width $b$,
$P_{crit}$ decreases exponentially with increasing $b$ in agreement 
with Fig. 2.

The behavior of $n(P)$ allows us to understand also the features of Fig. 1.
By approximating Eq.\ (\ref{SORE}a) with 
$n_2 \simeq G_2 \tau_e / (1 + n \lambda_2 \tau_e)$ we get
\begin{equation}
I_1 / I_2 = 
{\lambda_1 n_1 (n_1 -n) \over \lambda_2 n_2 (n_2 +n)}
\simeq {1 \over n \lambda_2 \tau_e }
\left[ 1 + {G_1 \over G_2} (1 + n \lambda_2 \tau_e) \right] .
\end{equation}
When $P \ll P_{crit}$ the electron tunneling time is almost constant,
$n \lambda_2 \tau_e \ll 1$ and the ratio $I_1 / I_2$ decreases
as $n^{-1}$, i.e. $P^{-1}$.
The behavior of $I_1 / I_2$  changes drastically at the critical power 
$P_{crit}$ due to the exponential change of $\tau_e$ and, finally,
for $n \lambda_2 \tau_e \gg 1$ the asymptotic value 
$I_1/I_2 = G_1/G_2$ is obtained.  

For purpouse of comparison with the following Section, 
in Fig. 3 we show the behavior of the calculated normalized photoluminescence 
intensities $I_i/I_{i \infty}$, i=1,2, in the same ADQW of Fig.s 1 and 2 
as a function of the barrier width $b$ and for different absorbed powers.
The normalization $I_{i \infty}$ is the photoluminescence current density 
for $b\to\infty$. 
The corresponding self-consistent electric field is shown in Fig. 4.
The normalized photoluminescence intensity of well 2 - Fig. 3 - vanishes 
at smaller $b$ when the tunneling current becomes higher. 
In this limit $J_e \to G_2$.
At the same time the normalized photoluminescence intensity of well 1 
tends to $(G_1+J_e)/G_1=1+G_2/G_1$.
The self-consistent electric field - Fig. 4 - gradually increases 
for decreasing $b$,
until, in the tunneling dominated limit $I_2/I_{2 \infty} \to 0$, it
vanishes exponentially.

\section{tunneling from a quantum well to surface states}

The considerations developed in the previous sections can be extended 
to describe the case where a single quantum well is close to 
a surface. This is another situation where tunneling followed by 
recombination takes place \cite{FE,2,3,4,japl}. 
Electron and hole bands of defect states localized at the surface
play the role of the $e1$ and $hh1$ bands of the missing well.
In a steady-state situation the system can be still described by
the rate equations (\ref{RE}) where the index 1 is associated 
with the surface and the index 2 with the well.
Let $n_1$ and $p_1$ be the charge concentrations in the 
donor-like and acceptor-like surface bands.
We can neglect the photogeneration of pairs at the surface 
so that $G_1=0$.
Moreover $\lambda_1 n_1 p_1 = I_1$ represents the nonradiative 
recombination current density at the surface. 
Since electrons and holes recombine (through a multiphonon process)
very fast with respect to the other relevant time scales, one 
is allowed to take $\lambda_1 \to \infty$.

The tunneling current densities from the occupied states of well 2
to the empty surface states 1 are still given by the general expressions 
(\ref{JEG}) and (\ref{JHG}).
Now, however, the surface densities $\nu_e^{(1)}$ and $\nu_h^{(1)}$ 
are not constant in general. 
This implies the following definition for the corresponding Fermi energies:
\begin{equation}
\int_{E_{e1}^{(1)}+eFb-E_{e1}^{(2)}}^{\epsilon_F^{(1)}} 
~\nu_e^{(1)} (\epsilon)~d\epsilon = n_1
\label{EFE}
\end{equation}
for the donor-like band, and
\begin{equation}
\int_{E_{hh1}^{(1)}-eFb-E_{hh1}^{(2)}}^{\epsilon_F^{(1)}} 
~\nu_h^{(1)} (\epsilon)~d\epsilon = p_1
\label{EFH}
\end{equation}
for the acceptor-like band.
As in the case of the two wells, we measure the energies from the bottom
of the $e1$ or $hh1$ band of well 2.
More importantly, $\nu_e^{(1)}$ and $\nu_h^{(1)}$
depend on the material characteristics 
and can be orders of magnitude different. 
If $\nu_e^{(1)} \ll \nu_h^{(1)}$, the effective mass 
difference between electrons and holes can be overcompensated by the difference
in the surface densities and holes can tunnel more effectively than
electrons do. This is the case, for instance, at the $GaAs$-oxide interface
\cite{japl,Spicer}. 
In this case holes accumulate at the surface and in a steady-state
situation $p_1 > n_1$. 
The electric field still given by Eq.\ (\ref{F}) turns out to be negative. 

As in the case of the two wells, the tunneling current densities
can be approximated by $J_e=n_2/\tau_e$ and $J_h=p_2/\tau_h$. 
Now, however, we distinguish two cases.
If $n_1>p_1$, the tunneling rates are given by Eq.\ (\ref{TAUE}) 
and Eq.\ (\ref{TAUH}) with $\nu_e^{(1)}=\nu_e^{(1)}(0)$ 
and $\nu_h^{(1)}=\nu_h^{(1)}(0)$. 
If $p_1>n_1$, we have 
\begin{equation}
{1 \over \tau_e} = {2 \pi \over \hbar} 
\left| \langle \Phi^{(1)}_0 |V_e| \Phi^{(2)}_0 \rangle \right|^2  
A~ \nu_e^{(1)}(0)
\label{TAUES}
\end{equation}
\begin{equation}
{1 \over \tau_h} = {2 \pi \over \hbar} 
\left| \langle \Phi^{(1)}_0 |V_h| \Phi^{(2)}_0 \rangle \right|^2  
A ~\nu_h^{(1)}(0)~
f\left({ \epsilon_F^{(1)} \over k_B T} \right) ~.
\label{TAUHS}
\end{equation}

When the electric field and the tunneling rates are known, the solution
of the initial rate equations is still given by Eq.\ (\ref{SORE}). 
With the conditions $G_1=0$ and $\lambda_1 \to \infty$ we get
the simpler formulas
\begin{mathletters}
\label{SORES}
\begin{equation}
n_2 = \sqrt{ \left({n\over2}+{1\over2\lambda_2 \tau_e} \right)^2 +
{G_2 \over \lambda_2} } -
\left({n\over2}+{1\over2\lambda_2 \tau_e} \right)
\end{equation}
\begin{equation}
n_1 = n
\end{equation}
\begin{equation}
p_1=0
\end{equation}
\begin{equation}
p_2=n_2+n ~.
\end{equation}
\end{mathletters}
The same recursive method explained in section II, 
allows us to find the steady-state values of the
electric field, of the charge concentrations and of the tunneling rates. 

The evaluation of the luminescence intensity $I_2$ implies the knowledge 
of the nature of the surface states. 
We can try to get information on the surface states by fitting 
experimental photoluminescence data.
We will concentrate on the specific example of an $Al_{0.3}Ga_{0.7}As$ 
surface with a nearby $GaAs$ quantum well \cite{FE,japl}.
At energy close to the bottom of the $e1$ band of well 2 the  
$Al_{0.3}Ga_{0.7}As$ surface has only donor-like states
belonging to the exponentially vanishing Urbach tail
\begin{equation}
\nu_e^{(1)}(\epsilon) = { m_e^{(1)} \over \pi \hbar^2 } 
\exp \left( -{\Delta E_c + eFb - E_{e1}^{(2)} - \epsilon 
\over \epsilon_e} \right)  ~.
\label{URBACH}
\end{equation}
Such states are assumed to be nodal hydrogenic wavefunctions 
\cite {L} with radius $r_e$ fixed by the depth into the gap.
Their explicit expression is given in Appendix B.
We assume that at the top of the gap the state density is the 
two dimensional density of free $Al_{0.3}Ga_{0.7}As$ electrons 
with effective mass $m_e^{(1)}$.
The parameter $\epsilon_e$ will be considered as a fitting parameter.
According to Eq.\ (\ref{URBACH}) and Eq.\ (\ref{EFE}) where we
can assume $E_{e1}^{(1)} = -\infty$,  
the Fermi energy for the donor-like surface band is
\begin{equation}
\epsilon_F^{(1)} =  \Delta E_c + eFb - E_{e1}^{(2)} + \epsilon_e 
\ln \left( {\pi \hbar^2 n_1 \over m_e^{(1)} \epsilon_e} \right) ~.
\end{equation}
On the other hand, at energy close to the bottom of the $hh1$ of well 2 
the $Al_{0.3}Ga_{0.7}As$ surface has a very high concentration of
acceptor-like defect states \cite{Spicer}. 
We schematize them again by nodal hydrogenic wavefunctions
\cite{L} but with radius $r_h$ to be considered as a second fitting 
parameter.
These states are assumed to be distributed in energy with constant 
density $\nu_h^{(1)}$.
The Fermi energy for the acceptor-like surface band is then
\begin{equation}
\epsilon_F^{(1)} = p_1/\nu_h^{(1)} + E_{hh1}^{(1)} + eFb - E_{hh1}^{(2)} ~.
\end{equation}
The tunneling matrix elements for the above surface states
and surface densities are evaluated explicitly in Appendix B.

According to the experiment reported in \cite{FE} we choose 
the well width $a_2=60$ \AA\, the temperature $T=4.2$ K,
the photon energy $h\nu=1.608$ eV and the incident power density 
$P_i=0.5$ Wcm$^{-2}$. 
The incident efficiency is estimated to be 1\% and we take
$G_2=0.01~P_i/h\nu$.
The relevant material parameters are \cite{AB}:
$\Delta E_c=0.3$ eV, $\Delta E_v=0.128$ eV, 
$m_e^{(1)}=0.091~m$, $m_e^{(2)}=0.067~m$, $m_h^{(2)}=0.34~m$, 
$m$ being the free electron mass, and $\varepsilon_r=12$. 
Moreover we put $\lambda_2=6$ cm$^2$ s$^{-1}$ \cite{SEKK}.
We assume an acceptor-like surface state density 
$\nu_h^{(1)}=10^{14}$ cm$^{-2}$eV$^{-1}$ \cite{rita} with 
$E_{hh1}^{(1)}\simeq 1$ eV
(the results we found do not depend crucially on this particular value).
The free parameters, $\epsilon_e$ and $r_h$, are fixed by 
fitting the normalized photoluminescence intensity $I_2/I_{2 \infty}$
to the experimental data \cite{FE} obtained for different values of 
the barrier width $b$. 
A least-square-error procedure gives the unique solution  
$\epsilon_e=12$ meV and $r_h=11$ \AA.
In Fig. 5, comparison is made between the ratio $I_2/I_{2 \infty}$, 
calculated with these values, and the experimental data. 

Due to presence of two parameters 
the agreement between theoretical and experimental data in Fig. 5
should be considered as a source of information for these two parameters
in the case the model is valid more than a proof of validity of the
model itself.
A reliable check of validity of our model is obtained by 
comparing the theoretical values of the self-consistently estimated 
electric field $F$ with 
the values deduced from measured Stark shifts as a function of the QW 
excitation \cite{japl}. 
The calculated field values are shown in Fig. 6 as a function of the barrier 
thickness $b$, for different values of the incident power. 
It is seen that, for high levels of excitation, the field approaches values 
of order $10^{5}$ V cm$^{-1}$ and
keeps increasing when the barrier becomes thinner. 
Direct comparison with experiment
is made for a 80-\AA-thick barrier, a choice dictated by the wide laser power 
range where data were available for this moderately tunneling sample.
To obtain the value of the electric field from the measured 
Stark shift $\Delta E_p$ 
we use the relationship 
$\Delta E_p = K F^{3/2}$, empirically deduced from known results in the 
literature \cite{Gobel,ibm}. 
The constant $K$ is fixed by imposing that the electric field obtained
from the measured Stark shift $\Delta E_p$ = 0.33 meV, corresponding to
an incident power density $P_i$ = 0.68 W cm$^{-2}$, coincides with the
calculated value. We obtain
$ K = 1.25 \cdot 10^{-7}$ meV (V cm$^{-1}$)$^{-3/2}$, 
which compares well with the
values extracted from the data of Ref. \cite{Gobel}, $ K \simeq 1
 \cdot 10^{-7}$ meV (V cm$^{-1}$)$^{-3/2}$,
and Ref. \cite{ibm}, $ K \simeq 0.6 \cdot 10^{-7}$ meV (V cm$^{-1}$)$^{-3/2}$.

The comparison between the calculated and experimentally deduced electric
field is shown in Fig. 7 as a function of the incident power density.
The agreement is very good for a power range extending over 3 orders of
magnitude. 
The near-coincidence of the scale factor $K$ with other independently
obtained values should not be attributed
much importance, on one side because of
the approximations used in the model, e.g. density and distribution of 
surface states, and, on the other side, because of the experimental 
Stark shift is observed in the luminescence from the quantum
well, where the field is non-uniform and possibly different from its value 
in the 
barrier \cite{japl}. Instead, the correct functional dependence of the field
on the excitation level is a clear confirmation that the self-consistent 
approach provides a reasonably accurate description of the 
whole process.

\section{conclusions}

The photoluminescence efficiency
in asymmetric double quantum wells and in near-surface quantum wells is
strongly influenced by the tunneling of both electrons and holes between
the two wells, or the well and the surface states. The theoretical model
presented here takes into account this effect and allows a quantitative 
prediction for the photoluminescence rates and the value of the electric 
field which needs to be established across the tunneling barrier in a 
steady-state situation.
Experimental results regarding both ADQW and near-surface QWs, 
namely changes 
in photoluminescence efficiency and 
peak shifts due to the self-induced electric field, are very 
well accounted for.

Our model can be readily generalized to include 
the case of an externally applied electric field 
as well as the case of not constant excitation.

%
%%%%%%%%%%%%%%%%%%%%%%%%%%%%%%%%% ACKNOWLEDGMENTS
%
\acknowledgments

We thank M. Capizzi, B. Bonanni and M. Colocci for very profitable 
discussions.
%
%%%%%%%%%%%%%%%%%%%%%%%%%%%%%%%%% APPENDIX
%
\appendix
\section{adqw matrix elements}

We call $z$ the coordinate orthogonal to the interfaces and
suppose that the quantum well 1 is in $-a_1 \leq z \leq 0$
and the quantum well 2 in $b \leq z \leq b+a_2$.
Firstly let us consider the case of electrons. 
As schematized in Fig. 8 we assume a rectangular potential profile with 
left and right discontinuities 
$V_l^{(1)}=\Delta E_c$ and $V_r^{(1)}=\Delta E_c-eFb/2$ for the well 1 and
$V_l^{(2)}=\Delta E_c+eFb/2$ and $V_r^{(2)}=\Delta E_c$ for the well 2 where
$F$ is the electric field in the barrier region $0 \leq z \leq b$.
The electron wavefunctions at energy $\epsilon$, measured from the bottom
of the $e1$ band of well 2, are 
\begin{equation}
\Phi^{(1)}_{\epsilon} (x,y,z) =
{e^{ik_x^{(1)}x+ik_y^{(1)}y} \over \sqrt{A} }  C^{(1)} 
\left\{  
\begin{array}{lll}
\sin\delta^{(1)} e^{k_l^{(1)}(z+a1)}~~~            & \mbox{$z<-a_1$}   \\
\sin[k^{(1)}(z+a_1)+\delta^{(1)}]~~~               & \mbox{$-a_1<z<0$} \\
\sin[k^{(1)}a_1+\delta^{(1)}] e^{-k_r^{(1)}z}~~~   & \mbox{$0<z$}
\end{array}  
\right.
\label{PHI1}
\end{equation}
where
\begin{equation}
{\hbar^2 \over 2m_e^{(1)}} [(k_x^{(1)})^2+(k_y^{(1)})^2]
= \epsilon+E_{e1}^{(2)}-E_{e1}^{(1)}-eFb
\end{equation}
\begin{equation}
k_l^{(1)} = \sqrt{{2m_e^{(1)}\over\hbar^2}(\Delta E_c - E_{e1}^{(1)})}
\end{equation}
\begin{equation}
k^{(1)} = \sqrt{{2m_e^{(1)}\over\hbar^2} E_{e1}^{(1)}}
\end{equation}
\begin{equation}
k_r^{(1)} = \sqrt{{2m_e^{(1)}\over\hbar^2}(\Delta E_c -eFb/2 - E_{e1}^{(1)})}
\end{equation}
for the well 1 and
\begin{equation}
\Phi^{(2)}_{\epsilon} (x,y,z) =
{e^{ik_x^{(2)}x+ik_y^{(2)}y} \over \sqrt{A} }  C^{(2)}
\left\{  
\begin{array}{lll}
\sin\delta^{(2)} e^{k_l^{(2)}(z-b)}~~~                & \mbox{$z<b$}   \\
\sin[k^{(2)}(z-b)+\delta^{(2)}]~~~                    & \mbox{$b<z<b+a_2$} \\
\sin[k^{(2)}a_2+\delta^{(2)}] e^{-k_r^{(2)}(z-b-a_2)}~~~ & \mbox{$b+a_2<z$}
\end{array}  
\right.
\label{PHI2}
\end{equation}
where
\begin{equation}
{\hbar^2 \over 2m_e^{(2)}} [(k_x^{(2)})^2+(k_y^{(2)})^2] = \epsilon
\end{equation}
\begin{equation}
k_l^{(2)} = \sqrt{{2m_e^{(2)}\over\hbar^2}(\Delta E_c + eFb/2 - E_{e1}^{(2)})}
\end{equation}
\begin{equation}
k^{(2)} = \sqrt{{2m_e^{(2)}\over\hbar^2} E_{e1}^{(2)}}
\end{equation}
\begin{equation}
k_r^{(2)} = \sqrt{{2m_e^{(2)}\over\hbar^2}(\Delta E_c - E_{e1}^{(2)})}
\end{equation}
for the well 2.
Energy $E_{e1}^{(i)}$ measures the bottom of the $e1$ band
in the well $i=1,2$ from the bottom of the same well and
is determined by solving
\begin{equation}
k^{(i)} a_i = \pi -
\sin^{-1} \left( {\hbar k^{(i)} \over \sqrt{2m_e^{(i)} V_l^{(i)}}} \right) -
\sin^{-1} \left( {\hbar k^{(i)} \over \sqrt{2m_e^{(i)} V_r^{(i)}}} \right) .
\end{equation}
The phase shifts are $\delta^{(i)}=\tan^{-1}(k^{(i)}/k_l^{(i)})$,
$i=1,2$.
The constants $C^{(i)}$, $i=1,2$, are fixed by normalizing 
the wavefunctions $\Phi^{(i)}_{\epsilon}$: 
\begin{equation}
C^{(i)}=
\left[ {\sin^2 \delta^{(i)} \over 2 k_l^{(i)}} +
{\sin^2 (k^{(i)}a_i + \delta^{(i)}) \over 2 k_r^{(i)}} +
{a_i \over 2} -
{\sin[2(k^{(i)}a_i + \delta^{(i)})] - \sin (2 \delta^{(i)})
\over 4 k^{(i)}} \right]^{-1/2} .
\end{equation}
For $\epsilon=0$ and with the assumption that $V_e$ in Eq.\ (\ref{TAUE})
vanishes everywhere but in the barrier region where $V_e=\Delta E_c$, 
the tunneling matrix element is
\begin{eqnarray}
& &\langle \Phi^{(1)}_0 | V_e| \Phi^{(2)}_0 \rangle = \nonumber \\
& &\int_0^{\sqrt{A}} dx \int_0^{\sqrt{A}} dy \int_0^b dz~
{e^{-ik_x^{(1)}x-ik_y^{(1)}y} \over \sqrt{A} }  C^{(1)} 
\sin[k^{(1)}a_1+\delta^{(1)}] e^{-k_r^{(1)}z}
~\Delta E_c~ 
{1 \over \sqrt{A} }  C^{(2)}
\sin\delta^{(2)} e^{k_l^{(2)}(z-b)}  \nonumber \\
& &= {C^{(1)}C^{(2)} \over A} 
~\sin[k^{(1)}a_1+\delta^{(1)}] \sin\delta^{(2)}
~{ e^{-k_r^{(1)}b} - e^{-k_l^{(2)}b} \over k_l^{(2)} - k_r^{(1)} } 
~{1-e^{-ik_x^{(1)}\sqrt{A}} \over ik_x^{(1)} }
~{1-e^{-ik_y^{(1)}\sqrt{A}} \over ik_y^{(1)} }  .
\end{eqnarray}
The last two factors in this expression quickly oscillates with
$k_x^{(1)}$ and $k_y^{(1)} (k_x^{(1)}) =\sqrt{2m_e^{(1)} 
(E_{e1}^{(2)}-E_{e1}^{(1)}-eFb)/\hbar^2 - (k_x^{(1)})^2 }$.
By taking the value for $k_x^{(1)}=0$ (maximum value)
the electron tunneling matrix element is estimated as 
\begin{equation}
\left| \langle \Phi^{(1)}_0 |V_e| \Phi^{(2)}_0 \rangle \right|  \simeq  
\left| {2\hbar \Delta E_c C^{(1)}C^{(2)} 
\sin[k^{(1)}a_1+\delta^{(1)}] \sin\delta^{(2)}
\over 
\sqrt{2m_e^{(1)} A (E_{e1}^{(2)}-E_{e1}^{(1)}-eFb)}} ~~
{ e^{-k_r^{(1)}b} - e^{-k_l^{(2)}b} \over k_l^{(2)} - k_r^{(1)} } 
\right|  .
\end{equation}

In the case of holes we have a completely analogous situation 
where the relevant band is $hh1$ instead of $e1$. The
expressions derived for electrons still hold for holes with the 
substitutions $m_e^{(i)} \to m_h^{(1)}$, $i=1,2$, 
$\Delta E_c \to \Delta E_v$, $eFb \to -eFb$.

\section{near-surface QW matrix elements}

Following the notations of Appendix A, the surface is defined 
by the plane $z=0$ and the well 2 is in $b<z<b+a_2$.
Firstly, we consider the case of electrons and we measure the energy
from the bottom of the $e1$ band of well 2. 
The state $\Phi^{(2)}_{\epsilon} (x,y,z)$ is given by Eq.\ (\ref{PHI2}).
The donor-like surface state $\Phi^{(1)}_{\epsilon} (x,y,z)$ is approximated by 
a truncated $2p$ hydrogenic wavefunction \cite{L}  
\begin{equation}
\Phi^{(1)}_{\epsilon} (x,y,z) =
{z \over 4 \sqrt{\pi} r_e^{5/2} } e^{r/2 r_e}
\left\{  
\begin{array}{ll}
0  ~~~   & \mbox{$z<0$} \\
1  ~~~   & \mbox{$0<z$} 
\end{array}  
\right.
\label{PHI1S}
\end{equation}
where $r=\sqrt{x^2+y^2+z^2}$.
The state is at energy $\hbar^2 / (8 m_e^{(1)} r_e^2)$ below
the bottom of the conduction band for the barrier material where 
the electron effective  mass is $m_e^{(1)}$.
By imposing this energy to correspond to $\epsilon$ we determine the
radius  $r_e$ 
\begin{equation}
r_e = {\hbar \over \sqrt{ 8 m_e^{(1)} 
\left( \Delta E_c - E_{e1}^{(2)} + eFb - \epsilon  \right) } }
\label{RAE}
\end{equation} 
Using parabolic coordinates $\eta=(r-z)/r_e$, $\xi=(r+z)/r_e$,
$\varphi=\tan^{-1}(y/x)$, the tunneling matrix element between the  
well and surface states at $\epsilon=0$ is: 
\begin{eqnarray}
& & \langle \Phi^{(1)}_0 |V_e| \Phi^{(2)}_0 \rangle =
\nonumber \\
& &\int_0^{\infty} d\eta \int_\eta^{\eta+2b/r_e} d\xi \int_0^{2\pi} d\varphi~
{r_e^3 (\xi+\eta) \over 4}~~
{(\xi-\eta) e^{- (\xi+\eta)/4} \over 8 \sqrt{\pi} r_e^{3/2}} 
~\Delta E_c~ 
{C^{(2)} \sin\delta^{(2)} e^{k_l^{(2)}[(\xi-\eta)r_e/2-b]}
\over \sqrt{A} }  = \nonumber \\
& & {\sqrt{\pi} r_e^{3/2} \Delta E_c C^{(2)} \sin\delta^{(2)} 
\over 8 \sqrt{A}} 
\left\{
e^{-k_l^{(2)}b} \left[ {1\over 4 (2k_l^{(2)}r_e-1)^2} 
- {1\over 32 (2k_l^{(2)}r_e-1)^3} \right] \right. + \nonumber \\
& & \left.
e^{-b/2r_e} \left[ {(b/r_e)^2 +2 b/r_e \over 2k_l^{(2)}r_e-1}- 
{1+b/r_e \over 4 (2k_l^{(2)}r_e-1)^2}+ 
{1\over 32 (2k_l^{(2)}r_e-1)^3} \right]  \right\}
\label{MEL}
\end{eqnarray}

In the case of holes we have a completely analogous situation 
where the relevant band of well 2 is $hh1$ instead of $e1$
and the acceptor-like surface state is given by Eq.\ (\ref{PHI1S})
with $r_e \to r_h$.
Equation (\ref{MEL}) gives the tunneling matrix element for holes 
with the substitutions $r_e \to r_h$, $\Delta E_c \to \Delta E_v$, 
$eFb \to -eFb$.

%
%%%%%%%%%%%%%%%%%%%%%%%%%%%%%%%%% REFERENCES LIST
%

%
%%%%%%%%%%%%%%%%%%%%%%%%%%%%%%%%% FIGURE CAPTION
%
\begin{figure}
\caption{Calculated emission intensity ratio $I_1/I_2$ from two coupled
asymmetric wells
{\it vs} absorbed power density $P$ for various values of the barrier width
$b$.}
\label{FIG1}
\end{figure}

\begin{figure}
\caption{Calculated electric field $F$ which establishes 
between the two wells, in the same cases of Fig. 1.}  
\label{FIG2}
\end{figure}  

\begin{figure}
\caption{Calculated normalized photoluminescence intensities 
$I_1/I_{1 \infty}$ and 
$I_2/I_{2 \infty}$ {\it vs} barrier thickness $b$ for different
absorbed power densities $P$. $I_{i \infty}$ is the photoluminescence 
intensity from well $i$ in the limit $b\to\infty$.}
\label{FIG3}
\end{figure}

\begin{figure}
\caption{Calculated electric field $F$ which establishes between
the two wells in the same cases of Fig. 3.} 
\label{FIG4}
\end{figure}    

\begin{figure}
\caption{Normalized photoluminescence 
ratio $I_2/I_{2 \infty}$ of a near-surface well {\it vs} the surface-barrier 
thickness $b$. Dots: experimental data from Ref. [1]; solid line: 
best fitting in terms of the self-consistent model. Incident power 
density is $P_i$ = 0.5 W cm$^{-2}$.}
\label{FIG5}
\end{figure}

\begin{figure}
\caption{Calculated  electric field $F$
across the surface barrier {\it vs} the 
surface barrier thickness $b$ for different incident power densities $P_i$.}
\label{FIG6}
\end{figure}

\begin{figure}
\caption{Comparison between the electric field $F$ calculated from the model 
(solid line) and deduced from the measured Stark shifts [8] (dots)
{\it vs} the incident power density $P_i$. The surface-barrier thickness 
is b = 80\AA. }
\label{FIG7}
\end{figure}   

\begin{figure}
\caption{Energy-space diagram for electrons 
in the asymmetric double quantum well.}
\label{FIG8}
\end{figure}   
%
%%%%%%%%%%%%%%%%%%%%%%%%%%%%%%%%% END
%
\end{document}